\documentclass[prb,twocolumn,a4paper,groupedaddress,superscriptaddress,showkeys,showpacs]{revtex4}
\usepackage{amssymb,amsmath,graphicx}
\begin{document}
\title[]{Enhanced dispersion interaction between quasi-one dimensional conducting collinear
structures}
\author{Angela White}
\email{Angela.White@anu.edu.au}
\affiliation{Nanoscale Science and Technology Centre, Griffith
University, Nathan, QLD 4111, Australia} \affiliation{Centre for
Gravitational Physics, Department of Physics, Faculty of Science,
The Australian National University, Canberra ACT 0200, Australia}
\author{John F. Dobson}
\email{j.dobson@griffith.edu.au}
\affiliation{Nanoscale Science and Technology Centre, Griffith
University, Nathan, QLD 4111, Australia}

\begin{abstract}
Recent investigations have highlighted the failure of a sum of
$R^{-6}$ terms to represent the dispersion interaction in parallel
metallic, anisotropic, linear or planar nanostructures [J.\ F.\
Dobson, A.\ White, and A.\ Rubio, Phys.\ Rev.\ Lett.\ \textbf{96},
073201 (2006) and references therein]. By applying a simple
coupled plasmon approach and using electron hydrodynamics, we
numerically evaluate the dispersion (non-contact van der Waals)
interaction between two conducting wires in a collinear pointing
configuration. This case is compared to that of two insulating
wires in an identical geometry, where the dispersion interaction
is modelled both within a pairwise summation framework, and by
adding a pinning potential to our theory leading to a standard
oscillator-type model of insulating dielectric behavior. Our
results provide a further example of enhanced dispersion
interaction between two conducting nanosystems compared to the
case of two insulating ones.  Unlike our previous work, this
calculation explores a region of relatively close coupling where,
although the electronic clouds do not overlap, we are still far
from the asymptotic region where a single power law describes the
dispersion energy. We find that strong differences in dispersion
attraction between metallic and semiconducting / insulating cases
persist into this non-asymptotic region. While our theory will
need to be supplemented with additional short-ranged terms when
the electronic clouds overlap, it does not suffer from the
short-distance divergence exhibited by purely asymptotic theories,
and gives a natural saturation of the dispersion energy as the
wires come into contact.
\end{abstract}

\maketitle

\section{Introduction}

The dispersion interaction constitutes the 
outer, attractive part of the van der Waals (vdW) force (in the
non--retarded regime, 
for charge-neutral, non-polar species). It can be explained via
the interaction between small instantaneous dipoles arising due to
a mutual dynamic polarization of the electron clouds (see e.g.
[1]). These attractive forces, although weaker in magnitude than
ionic or covalent bonds between atoms or molecules, are ubiquitous
across nature and can play a central role in soft matter and
biophysical systems. Examples range from protein
folding\cite{proteinvdw1,proteinvdw2,proteinvdw3} to the adhesive
properties of gecko feet\cite{geko1,geko2}. 
While the present work cannot describe the contact regime where
the van der Waals force is strongest, it does suggest that
noncontacting regions may contribute more than previously
suspected to the energetics, in cases where pi-conjugation, for
example, may lead to near-metallic behavior in low-dimensional
structures.

Dispersion forces also play an important role in the rapidly
progressing area of nanoscience. Nanotubes are quasi--one
dimensional structures, with electronic properties determined by
their composition. The metallic or semi--conducting behavior
depends on the helicity of carbon nanotubes. Nanotubes composed of
boron nitride are primarily semi--conducting, with a wide band gap
and stable electronic properties \cite
{RCC94BNnano,MWMORR04CBNnanot}. Armchair (n,n) carbon nanotubes
are metallic. The vdW attraction facilitates the self--assembly of
single--walled nanotubes into bundles or
ropes\cite{BBBMR02revcnano} and being the primary inter--tube
attraction is important in a description of
the cohesive energetics of nanotube bundles\cite{D99carcnano}. 

Recently the differing asymptotic behavior of vdW interactions for
conducting systems, compared to insulating systems with identical
geometry, has been displayed for various systems \cite
{JDARvdw05,JDAWARvdwprl06,JFDAstJChem01,JFDIntJChem05}. The
simplest standard techniques applied in computing the vdW
interaction are based on the pairwise summation of all $R^{-6}$
contributions \cite{MNDispForces} between microscopic elements
separated by distance R, and so inherently assume local properties
of the two materials. More sophisticated techniques
\cite{LifshitzMolAttrFrcsSols,lifshitzGenThevdwFrcs} when coupled
with the usual assumption of a local bulk--like dielectric
function within the boundaries of the constituents, still lead to
an interaction asymptotically
equivalent to a sum of $R^{-6}$ contributions. 
Recently, progress has been made in formalisms that describe both
the contact region of electron cloud overlap, as well as a
dispersion contribution at large distances
\cite{vdWFnalGenGeomDionPRL04,softmattRydberg03,vdWRydberg03PRL,vdWdfHult96PRL,dfRydberg00,vdWSchroder03,benzPudzer06,vdWdftChakarova06,vdWdftLangreth05}.
However these theories predict the usual asymptotic $\Sigma
R^{-6}$ behavior and so could still be improved with respect to
description of low-dimensional systems with a small or zero
electronic energy gap.

For a determination of the vdW interaction between highly
anisotropic systems of metallic nature, a pairwise summation
method fails to incorporate the long wavelength, incompletely
screened electron density fluctuations. These lead to unexpected
power laws for the vdW interaction as a function of separation
\cite{JDAWARvdwprl06}. The effects of this physics appear in
calculations on parallel 1D electron gases
\cite{vdWParallelWiresChainsChang1971}, on two dimensional
electron gases \cite{SB98QuWells,DWPRL99,BS00MeFilms}, and in
graphene sheets \cite{JDAWARvdwprl06}. More recently this has been
highlighted as a more general phenomenon \cite{JDAWARvdwprl06}.
The greatest underestimation of the vdW interaction occurs between
distant parallel quasi one--dimensional conductors
\cite{vdWParallelWiresChainsChang1971,ASTparallel,JDAWARvdwprl06}.

The coupled plasmon approach addresses the shortfalls of the
pairwise summation technique, incorporating polarization between
multiplets of atoms in the system and also electron movement
within the wires. In modelling the electronic motions as coupled
plasmons, large electron displacement across many atoms in a
system is allowed to occur, providing a less constrained
representation of electron movement. Plasmon frequencies can be
determined by application of electron hydrodynamics and the vdW
dispersion energy then corresponds to the separation dependent
part of the sum of zero point plasmon energies. This approach has
been employed previously in determination of the vdW interaction
between systems of thin parallel metallic plates and also thin
parallel metallic wires. The sum of plasmon energies is also known
to provide an approximation to the full correlation energy in the
Random Phase Approximation (RPA). (See e.g. [13]). A full RPA
energy calculation would thus provide a natural and seamless
extension to the present method, valid at all separations
including the regime of full electronic cloud overlap.

For two thin parallel metallic plates of infinite area, separated
by a distance greater than the thickness of the plates, B\"ostrom
and Sernelius \cite{BS00MeFilms}, and Dobson and Wang
\cite{DWPRL99} have applied the coupled plasmon approach to obtain
a dispersion energy of dependence $D^{-5/2}$ on the separation
distance D of the metal slabs. This result is in agrement with the
appropriate limit of microscopic random phase approximation
calculations for a pair of two dimensional electron gases
performed by Sernelius and Bj\"ork \cite{SB98QuWells}. A pairwise
summation analysis for this system of thin parallel metallic
plates gives a $D^{-4}$ dependence of the vdW attraction, notably
smaller at large separation than the $D^{-5/2}$ dependence derived
by applying the coupled plasmon approach. An equally
unconventional result was obtained for the interaction between
graphene planes \cite{JDARvdw05,JDAWARvdwprl06}.

Another example system in which the pairwise summation analysis
was shown to underestimate the dispersion interaction for
conductors is that of two parallel wires of infinite length
\cite{vdWParallelWiresChainsChang1971,ASTparallel,JDAWARvdwprl06,honoursthesis}.
For a pair of parallel, conducting, infinitely long wires, the
dispersion interaction, calculated by a standard sum of $R^{-6}$
contributions between microscopic elements separated by distance
$R$, has a $D^{-5}$ dependence on the separation distance, $D$, of
the two wires. However an analysis of the zero--point energy of
the delocalized coupled one--dimensional plasmon modes parallel to
the long axis, modelled for wires of length $L>>D$, finds the vdW
interaction to have a
\begin{equation}
D^{-2}\left( ln(D/A)^{-3/2}\right)  \label{parallelwires}
\end{equation}
dependence on separation distance, $D$.  $A$ denotes the smearing
radius of a wire, representing the finite extent of the electronic
wave--function or electronic density fluctuation on the wire in
the direction perpendicular to the long axis. The analysis assumes
an electron mean free path greater than the separation distance of
the wires, which can be satisfied by bismuth
nanowires\cite{GC04bismuthnano} and conducting
nanotubes\cite{D99carcnano,SDD98carbnano}.  The vdW interaction
(\ref{parallelwires}) is almost three powers of $D$ greater than
that obtained in a standard pairwise summation framework.  This
highlights the importance of including electron movement and
screening from subsequent polarization of electron pairs in
conducting systems. The occurrence of an enhanced dispersion
interaction in metallic systems might have repercussions,
particularly in nanotechnology.  The above considerations suggest
metallic and non--metallic nanotubes could experience different
cohesive forces, for example though analysis beyond the present
methods would be needed to explore this possibility in the case of
electron cloud overlap.  In the present paper we explore another
case \cite{honoursthesis} where a sum of $R^{-6}$ contributions
does not yield a correct description of the vdW interactions.  We
consider two linear structures (``wires''), each of length $L$, in
a collinear ``pointing'' configuration as shown in figure 1.  $D$
is the separation between the near ends of the wires, while
$D_{cm}$ is the distance between the centers of the wires.  $A$ is
an effective radius of the wire, discussed further below.

Our motivation for looking at this case was threefold.  Firstly,
there is intrinsic interest in the cohesive properties of
nanostructures of all kinds, and the present work is particularly
relevant to the interactions of carbon nanotubes, a
technologically important case.  Secondly, recent work
\cite{Blundellphd} has suggested that van der Waals forces may be
involved in the surprisingly strong force that tends to make the
tips of iron micro--whiskers grow towards one another during
fluidized--bed iron ore reduction processes.  While it is not
clear that the electron mean free path in these cases is
sufficiently long to validate the assumptions of the present work,
it is interesting that our approach can predict vdW forces between
linear conducting systems in the appropriate ``pointing''
geometry, that are enhanced compared with standard vdW theory.
 Thirdly, a numerical investigation of wires of finite length
allows us to look at a region of relatively close coupling where,
although the electronic clouds are not permitted to overlap, we
are still far from the truly asymptotic regime where a single
power law describes the force.  We find that the dispersion
interaction between the wires shows strong differences between
metallic and insulating cases even in this non-asymptotic regime,
reinforcing asymptotic results
\cite{SB98QuWells,BS00MeFilms,vdWParallelWiresChainsChang1971,
JDARvdw05,JDAWARvdwprl06} already known in other geometries.  We
further find that the greater interaction at large distances in
the metallic case does \emph{not} imply a lesser interaction at
small distances in the metallic case, contrary to a false
expectation based on a single power law at all distances.

The paper is organized as follows.  The analytic work will be
presented in section \ref{analytics}, and section \ref{numerics}
then outlines the numerics.  Our results and discussion compare
the dispersion interaction between wires in the ``pointing''
geometry for insulators and conductors and are presented in
section \ref{resultsanddisc}.  The principal findings and future
prospects are discussed in the final section.

\section{Analytics}

\label{analytics}

\begin{figure}[tbp]
\begin{center}
\includegraphics[width=8cm]{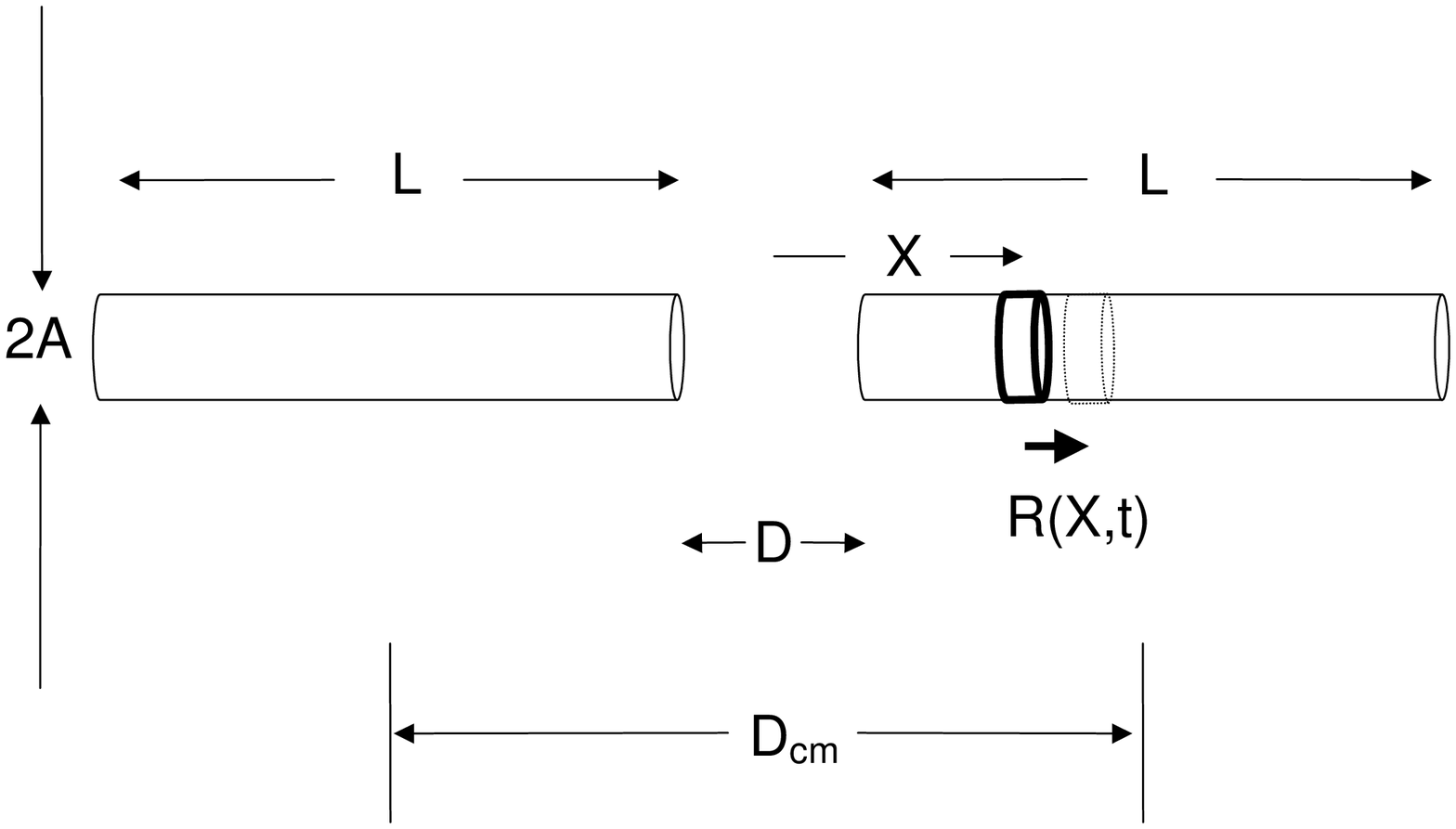}
\end{center}
\caption{}
\label{Fig1}
\end{figure}

We evaluate the dispersion interaction between the two wires by
the coupled--plasmon approach (see e.g.\ Ref.\ [12],[13],[14]).
 Plasmons are quantized versions of the wavelike collective motions
of the electrons.  The equilibrium /ground state of the wire is
described by a constant electron number density per unit length,
$n_0$.  Plasma waves entail a perturbation $\delta n(X,t)$ to the
number density, where we have defined a variable X (see Fig.\
\ref{Fig1}) to label positions along the wire.

Here we neglect electronic radial and azimuthal motions -- i.e.\
those perpendicular to the long axis.  These are frozen out by
quantum effects in atomically thin systems, but can be present in
other cases.  We do not consider them here because, for the thin
linear systems of interest here, they consist of spatially
constrained electronic motions leading to conventional van der
Waals force laws.  By contrast the unconstrained electronic
motions along the wire will be shown below to lead to an
unconventional, enhanced long ranged van der Waals interaction
that is the principal focus of attention here.  For example, the
azimuthal electronic motions have been considered by Rotkin and
Hess \cite{RH02nt}, in the case of two long parallel nanotubes.
 They considered only small separations between the tubes and found
these modes to give a more rapidly decaying van der Waals
interaction as a function of nanotube separation $D$, compared to
the purely longitudinal modes investigated in [11].
 Thus we expect that our neglect of the electronic motions
perpendicular to the long axis will not affect the long--ranged
forces that we are investigating here.

To describe the electronic motions along the wire we use a
conventional hydrodynamical picture, valid even for degenerate
quantal electrons in the limit of long--wavelength perturbations.
 During the plasmon motion, the electron fluid element that was at
position $X$ (see Fig.\ \ref{Fig1}) in the unperturbed state of
the wire (bold lines in Fig.\ 1) is displaced to position
$X+R(X,t)$ (dotted lines in Fig.\ \ref{Fig1}).  In the simplest
hydrodynamical model (correct to lowest order in the wavenumber
$Q$ of the wave) the motion of the fluid element is described by
Newton's second law for a a free mass under the action of a mean
potential energy function $\Phi(X,t)$ generated by the Coulomb
interaction with other fluid elements: $ Md^2R(X,t)/dt^2=-\partial
\Phi (X,t)/dX=F$.  Here $M$ is the electron effective mass for
motion along the wire.

An element of electron fluid at a point on any wire will
experience a Coulomb potential due to electrons at every other
point on that wire and also due to those at every point on the
adjacent collinear wire. Considering plasmons in one dimension,
evidently the Fourier transform of the Coulomb potential does not
converge. However, by recognizing that the one electron
wavefunctions are of finite extent in the direction perpendicular
to the wire, we can examine a symmetrically `smeared' version of
the wire with finite smearing radius, $A$.  While we apply a
radially smeared pair potential, expressed explicitly as
\begin{equation}  \label{cp}
\tilde{\phi}(\vert X-X^{\prime}\vert)= \frac{e^{2}}{\left(
(X-X^{\prime})^{2}+A^{2} \right) ^{1/2}}
\end{equation}
in our work to follow, any expression that saturates the Coulomb
interaction for $\vert r \vert<A$ would give similar results for
the long-wavelength fluctuation phenomena that drive the unusual
effects to be explored here.

The force on an element of electron fluid at a point in time can be
expressed using the radially smeared pair potential, $\tilde{\phi}$, as:
\begin{equation}
\begin{split}
M& \frac{d^{2}R(X,t)}{dt^{2}}=-\frac{\partial }{\partial X}\left(
\int\limits_{-D/2-L}^{-D/2}\delta n(X^{\prime },t)\tilde{\phi}(|X-X^{\prime
}|)dX^{\prime }\right. \\
{}& +\left. \int\limits_{D/2}^{D/2+L}\delta n(X^{\prime },t)\tilde{\phi}%
(|X-X^{\prime }|)dX^{\prime }\right)\,.
\end{split}
\label{NewtonII}
\end{equation}
This applies for either $D/2\leqslant X\leqslant D/2+L$ or
$-(D/2+L)\leqslant X\leqslant -D/2$.  We apply a linearized form
of the continuity equation,
\begin{equation}
\delta n(X,t)=-\frac{\partial }{\partial X}\left( n_{o}(X)R(X,t)\right) ,
\label{lincontinuity}
\end{equation}
to describe the perturbation to the electron number density in terms of the
equilibrium number of electrons per unit length of wire, $n_{o}(X)$.

Plasmon movement within metallic wires is confined by requiring
zero electron displacement at the wire ends.  An alternative
analysis, in which the zero boundary condition occurring at the
wire ends is naturally inherent, could be obtained by applying a
sine--basis decomposition.  Note that different boundary
conditions, such as allowing the electron gas to move out over the
ends of the uniform positive background, which would then cause a
restoring force, could be applied instead, and possibly would
provide a further alternative description of the plasmon modes.

We seek time--periodic separable solutions of (\ref{NewtonII}) in
the form
\begin{equation}  \label{XX}
R(X,t)=R(X)\exp(-i\Omega t)\,.
\end{equation}

The left--right symmetry evident in Fig.\ \ref{Fig1} requires that
there exist even solutions for which $R(X)=R(-X)$ and also odd
solutions for which $R(X)=-R(-X)$.  We used this property and put
(\ref{XX}) and (\ref{lincontinuity}) into (\ref{NewtonII}).
 Integrating by parts and using the explicit form of the smeared
Coulomb potential (\ref{cp}) we obtained

\begin{equation}
\begin{split}
\frac{M\Omega ^{2}R(X)}{n_{0}e^{2}}& =\int_{D/2}^{L+D/2}R(X^{\prime })\left(
\frac{-2(X-X^{\prime })^{2}+A^{2}}{\left( (X-X^{\prime })^{2}+A^{2}\right)
^{5/2}}\right. \\
& \left. \pm \frac{-2(X+X^{\prime })^{2}+A^{2}}{\left(
(X+X^{\prime })^{2}+A^{2}\right) ^{5/2}}\right) dX^{\prime }\,.
\end{split}
\label{DimensionedEqMotionNoPressNoPin}
\end{equation}

Eq.\ (\ref{DimensionedEqMotionNoPressNoPin}) holds for either
$-(L+D/2)\le X\le -D/2\,$ or $ D/2\le X\le L+D/2$, and is an
eigenvalue equation for the frequencies $\Omega_i$ of
self-sustaining plasma oscillations, (plasmons) corresponding to
coupled charge density fluctuations on the two wires.

\subsection{Evaluation of the dispersion energy}

The dispersion energy $E_{vdW}$ is then the part of the total
plasmon zero-point energy that depends on the separation $D$
between the wires. Thus

\begin{equation}
E_{vdW}(D)=\frac{\hbar }{2}\sum_{j}\left( \Omega _{j}(D)-\Omega
_{j}(D\rightarrow \infty )\right) \, . \label{DimensionedEDisp}
\end{equation}

Application of numerical techniques is necessary to solve for the
eigenfrequencies $\Omega _{i}\,$, as the broken translational
symmetry of the collinear wire geometry with a gap prevents
Fourier transformation being used to solve the integral equation
(\ref{DimensionedEqMotionNoPressNoPin}) analytically in k--space.

\subsection{Incorporating Internal Pressure}

Although the eigenvalue problem
(\ref{DimensionedEqMotionNoPressNoPin}) derived from the
expression for the force felt on an element of electron fluid at a
point could now be solved to give us odd and even solutions for
the plasmon frequency, we extend this to incorporate the
contribution to the force felt by an element of electron fluid
from the internal pressure of the electron fluid.  This electronic
pressure is caused mainly by velocity deviation of electrons in a
fluid element from the average velocity.  It is the electron
pressure term that distinguishes our formalism from a purely
classical Newton II approach, and it introduces the quantal and
Pauli-principle physics (electron degeneracy pressure) into the
formalism.  We expect it should give higher frequencies to the
plasmon modes with rapid spatial variation.  Without the pressure
term the theory correctly describes the response of a free
electron gas in the limit of long wavelength, and it agrees with
the corresponding limit of the quantal Lindhard response.  This
limit was adequate in [12] because there the treatment was limited
to the asymptotic regime of distant interactions between
infinitely long wires, where the long-wavelength fluctuations
completely dominate.  Here we want to explore wires of finite
length at finite separations.  While the new phenomena which we
expose will still turn out to be due to long-wavelength
fluctuations, a detailed treatment needs to account, for example,
for the reduced polarizability of a short wire of length $L$
resulting from the requirement to excite wavelengths of
$O(L^{-1})$ in order to polarize its electron gas. For this reason
we introduce the plasma pressure, in the simplest available model.
In some regimes it also aids in achieving numerical stability of
the solutions.

The excess pressure due to a density perturbation $\delta \rho$ is
usually modelled from an analysis of the free Fermi gas
\cite{FS78dp,E79dr,JFD02,DQ79hd,B70,B79hd} as $\delta
P=MB^{2}\delta \rho $.  Here $\delta \rho$ is the three
dimensional density perturbation and $B$ is a velocity of the
order of the Fermi velocity of the metal.  We take $\rho =
nA^{-2}$.  The pressure can now be expressed in terms of the
equilibrium pressure $P_{0}$, and the perturbation to the electron
number density per unit length, $\delta n = \delta \rho A^{2}$, as
$P=P_{0}+A^{-2}M B^{2}\delta n$.  Now $B$ is of order of the Fermi
velocity of the metal composing a wire.  The pressure force per
electron, to be added to the right side of (\ref{NewtonII}), is
then
\begin{equation}\label{pressforce}
F=-\frac{1}{\rho_{0}}\frac{dP}{dX}=MB^{2}\frac{d^{2}R(X,t)}{dX^{2
}}\,
\end{equation}
where from the linearized continuity (Eq (\ref{lincontinuity})),
the perturbation to the electron number density has again been
expressed as a function of the equilibrium number density, $n_{0}$
and the displacement, $R$ of an electron fluid element.

\subsection{Dispersion Interaction for Insulating wires}

The prior analysis, applying a coupled plasmon approach, accounts
for conduction of electrons and so provides a more complete
description in evaluating the vdW interaction between two
conducting collinear wires. However, we would expect the pairwise
summation method of individual $R^{-6}$ atomic contributions,
which does not allow for electron movement along the wires, to
remain an apt description of the dispersion interaction between
insulating wires with a sufficiently large gap. We therefore
applied two different modifications to the theory described above:
each is designed to explore the difference between metallic and
non-metallic wires.

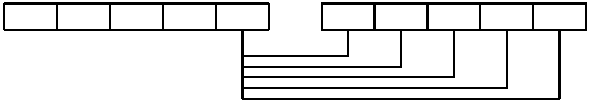
\begin{figure}[tbp]
\begin{picture}(500,50)
\put(0,30){\line(4,0){100}}\put(0,40){\line(4,0){100}}
\put(0,30){\line(0,1){10}}\put(20,30){\line(0,1){10}}
\put(40,30){\line(0,1){10}}\put(60,30){\line(0,1){10}}
\put(80,30){\line(0,1){10}}\put(100,30){\line(0,1){10}}
\put(120,30){\line(4,0){100}}\put(120,40){\line(4,0){100}}
\put(120,30){\line(0,1){10}}\put(140,30){\line(0,1){10}}
\put(160,30){\line(0,1){10}}\put(180,30){\line(0,1){10}}
\put(200,30){\line(0,1){10}}\put(220,30){\line(0,1){10}}
\put(90,30){\line(0,-1){26}}\put(90,20){\line(4,0){40}}
\put(130,20){\line(0,1){10}}\put(90,16){\line(4,0){60}}
\put(150,16){\line(0,1){14}}\put(90,12){\line(4,0){80}}
\put(170,12){\line(0,1){18}}\put(90,8){\line(4,0){100}}
\put(190,8){\line(0,1){22}}\put(90,4){\line(4,0){120}}
\put(210,4){\line(0,1){26}}
\end{picture}
\caption{Pairwise summation of an element of wire A with all
corresponding elements on wire B } \label{figurepw}
\end{figure}

(A) A modified pairwise additive approach was applied to calculate
the dispersion interaction between two collinear insulating wires.
Using (\ref{DimensionedEqMotionNoPressNoPin}), (\ref{pressforce})
and (\ref{DimensionedEDisp}), we first calculated numerically and
tabulated the dispersion interaction energy $E_{vdW}(\ell ,d)$ of
short metallic wires of length $L=\ell A$ with separation $D=dA$
between the ends. (For fixed $\ell <<d$ we found $E_{vdW}(\ell
,d)$ to be of the conventional form $-C(\ell )d^{-6}$ as
expected:\ see the Results section below).  We then modelled a
long \emph{insulating} wire of length $L=i\ell A,$ $(i=1,2,3,...)$
by a collection of short wire segments of equal length $\ell A$,
placed end to end, but with electrons not allowed to flow from
segment to segment.  This crudely represents individual atoms or
bonds on an insulating wire, as electron movement along the wire
is restricted, now being confined to within each wire segment.
 Assuming the mutual polarization between any two wire segments is
not affected by any other wire segment composing the collinear
wire system, the pre-tabulated dispersion interaction
$E_{vdW}(\ell ,d=q\ell )$ can be used between any two segments of
length $\ell A$ lying in opposite wires, and separated by distance
$q\ell $.  Summation of the dispersion interaction between each
segment on one wire and all segments on the opposite wire (as
illustrated in Fig.\ \ref{figurepw}) then gives the total vdW
interaction for insulating wires, separated by distance $D=p\,\ell
$, in the pairwise summation approach:
\begin{equation}\label{PairsummedInteraction}
E^{pairwise}_{vdW}(D=p\,\ell )=\sum_{m,n=0}^{i-1}E_{vdW}(\ell
,d=(p+m+n)\ell)\,.
\end{equation}
Here each wire has length $L=i\ell $ where $i$ is an integer.

This method provides an increasingly accurate representation of two
collinear wire insulators with decreasing length of the individual wire
segments, as electron movement is restricted to the individual wire pieces.

(B) We also explored a more conventional approach to model an
insulator, by adding a harmonic pinning force $-M\Omega
_{pin}^{2}R(X)$ to the right-hand side of the equation of motion
(\ref{NewtonII}) for $M\ddot{R}$. This, plus transposition and
inclusion of the pressure term (\ref{pressforce}) resulted in an
eigenvalue problem of the form (cf
(\ref{DimensionedEqMotionNoPressNoPin}))

\begin{equation}\begin{split}
& \frac{M\left( \Omega ^{2}-\Omega _{pin}^{2}\right)
}{n_{0}e^{2}}R(X) = \int_{D/2}^{L+D/2}R(X^{\prime })\times
 \\
& \left( \frac{-2(X-X^{\prime })^{2}+A^{2}}{ \left( (X-X^{\prime
})^{2}+A^{2}\right) ^{5/2}} \pm \frac{-2(X+X^{\prime
})^{2}+A^{2}}{\left( (X+X^{\prime })^{2}+A^{2}\right)
^{5/2}}\right) dX^{\prime }
 \\
&\quad \quad \quad-\frac{MB^{2}}{n_{0}e^{2}} \frac{d^{2}R}{dX^{2}}
\,. \label{DimensionedPinnedPressureEigvalEqu}
\end{split}\end{equation}

Each eigenfrequency of (\ref{DimensionedPinnedPressureEigvalEqu}) is of the
form

\begin{equation}
\Omega _{i}=\sqrt{\bar{\Omega}_{i}^{2}+\Omega _{pin}^{2}}
\label{PinnedFrequencies}
\end{equation}
where $\bar{\Omega}_{i}$ is an eigenfrequency of
(\ref{DimensionedPinnedPressureEigvalEqu}) with the pinning term
absent. A semiconducting or insulating wire with an electronic
energy gap $E_{g}$ can be modelled by including a finite pinning
frequency $\Omega _{pin}\neq 0$. A metallic wire is modelled by
setting $\Omega _{pin}=0.$  The idea is that each electron
experiences a force tending to tie it to its labelled equilibrium
position $X$. In this simple approach the pinning energy $\hbar
\Omega _{pin} $ is loosely associated with the electronic gap
$E_{g}$ of an insulator, $\hbar \Omega _{pin}\approx E_{g}$. The
pinning has the effect of reducing the polarizability of the
electron gas, stiffening the plasmon modes and making them less
affected by coupling between the wires. \ The modified frequencies
(\ref{PinnedFrequencies}) are then substituted into
(\ref{DimensionedEDisp}) to obtain the vdW energy as before. A
similar pinning approach gives the conventional power laws for the
vdW attraction between insulating two--dimensional sheets
($E_{vdW}\propto -D^{-4}$) and for distant parallel insulating
wires\cite{DobsonSurfSciInPress07} ($E_{vdW}\propto -D^{-5}$) . As
we will show in the following Results section, for the present
problem in the appropriate parameter regime the pinned approach
also produced results very similar to approach (A) in which the
wire was cut into short segments. Both will be seen to give
results for the van der Waals interaction between collinear wires
that is very different from the case of metallic wires.

\section{Numerics}

\label{numerics}

To solve Eq (\ref{DimensionedPinnedPressureEigvalEqu})
numerically, the integral is discretized by application of the
trapezoidal rule, which along with a zero boundary condition for
the displacement at the edges of each wire, converts the integral
equation (\ref{DimensionedPinnedPressureEigvalEqu}) to a symmetric
matrix eigenvalue equation of dimension $N-1$, where $N$ is the
total number of partitions of the integral term in Eq
(\ref{DimensionedPinnedPressureEigvalEqu}).  With inclusion of a
pinning force, the matrix eigenvalue problem is of the form
\begin{equation}
\frac{M(\Omega ^{2}-\Omega _{pin}^{2})}{n_{0}e^{2}}R_{i}=
\sum_{m=1}^{N-1}\left(G_{im}W_{m}-\frac{MB^{2}N^{2}}{L^{2}n_{0}e^{2}}P_{im}\right)R_{m}
 \label{Eqnmot}
\end{equation}
for $1 \le i \le N-1$. Here $G_{im}$ is the matrix
\begin{eqnarray*}
G_{im} &=&\frac{A^2-2\left( (i-m)L/N\right) ^2}{\left(
((i-m)L/N)^2+A^2\right) ^{5/2}} \\
&&\pm \frac{A^2-2\left( D+(i+m)L/N\right) ^2}{\left(
(D+(i+m)L/N)^2+A^2\right) ^{5/2}} \,.
\end{eqnarray*}
Here $W_m = L/N$, $1 \le m \le N-1$, is the weighting function for
trapezoidal integration: the end--weights $W_0 = W_N = L/2N$ are
not sampled because of the pinning condition $R_0 = R_N = 0$, so
that the $(N-1) \times (N-1)$ matrix equation (\ref{Eqnmot}) is
symmetric.

In (\ref{Eqnmot})
\begin{equation}
P_{im}=\delta _{i-1,m}-2\delta _{i,m}+\delta _{i+1,m}  \label{Pim}
\end{equation}
is a discrete dimensionless version of the second--derivative
operator.  In obtaining (\ref{Eqnmot}) from
(\ref{DimensionedPinnedPressureEigvalEqu}), $X$ and $X'$ have been
replaced by discrete positions $X_i$ and $X_m$ defined in terms of
the absolute wire length L, wire separation distance D and N, by
$X_i=D/2+iL/N$ for $i=0,1,2....N$ and $X_m=D/2+mL/N$ for
$m=0,1,2....N$.

\subsection{Dimensionless form of equations}

The eigenvalue problem can be written in dimensionless form as
follows. Firstly we scale all lengths by the effective radius $A$,
introducing dimensionless quantities $d,\,\ell ,$ $r_i$ and
$w_i\,$such that $D=dA$, $L=\ell A$, $\,R_i=r_iA$ and $W_i=w_iA$.
We also define $J_{im}=A^3G_{im}$ which is dimensionless. We
introduce a characteristic frequency $\Omega _0$ defined by $\,$
\begin{equation}\label{Omrga0}
\Omega _0=\sqrt{\frac{n_0e^2}{MA^2}}
\end{equation}
which is $\frac 12\Omega _P,$ where $\Omega _P=\sqrt{4\pi n_0(\pi
A^2)^{-1}e^2/M}\,$ is the bulk plasma frequency calculated
assuming a bulk electron density $n_{3D}=n_0(\pi A^2)^{-1}$, as
though the 1D density $n_0$ were distributed uniformly throughout
a cylinder of radius $A$. Then the equation of motion
(\ref{Eqnmot}) can be written in dimensionless form as
\begin{equation}
\frac{\Omega ^2-\Omega _{pin}^2}{\Omega _0^2}r_i=
\sum_{m=1}^{N-1}\left(J_{im}w_m-\beta ^2\frac{N^2}{\ell
^2}P_{im}\right)r_m \label{DimlessEquMotion}
\end{equation}
where
\begin{equation}
\begin{split}
J_{im}& =\frac{1-2\ell ^2(i-m)^2N^{-2}}{\left( \ell ^2(i-m)^2N^{-2}+1\right)
^{5/2}} \\
& \pm \frac{1-2\left( d+\ell (i-m)N^{-1}\right) ^2}{\left( \left( d+\ell
(i-m)N^{-1}\right) ^2+1\right) ^{5/2}}
\end{split}
\end{equation}
for even (+) and odd (-) solutions respectively. In
(\ref{DimlessEquMotion} ), \begin{equation} \beta
^2=\frac{B^2}{A^2\Omega _0^2}\,. \label{BetaSquared}
\end{equation}
is a dimensionless measure of the importance of the pressure term.

Equation (\ref{DimlessEquMotion}) holds for $1 \le i \le N-1$.

The discrete eigenfrequencies $\Omega _{j}$ from numerical
solution of the $ (N-1)\times (N-1)$ matrix equation (15) were
used as follows in calculating the attractive dispersion energy
$E_{vdW}$ between the wires (c.f. (7)):
\begin{equation}
E_{vdW}=\frac{\hbar }{2}\sum_{j=1}^{N-1}\left( \Omega
_{j}^{(+)}(d)+\Omega _{j}^{(-)}(d)-2\Omega _{j}^{(0)}(d\rightarrow
\infty )\right) . \label{EvdWEvenOdd}
\end{equation}%
Here $\Omega _{j}^{(+)}(d)$ is the $j^{th}$ eigenvalue of (15)
with the $+$ sign used in (16), corresponding to even plasmon
modes. \ Similarly \ $ \Omega _{j}^{(-)}(d)$ is the eigenvalue
with the $-$ sign in (15), corresponding to odd modes. \ $\Omega
_{j}^{(0)}(d\rightarrow \infty )$ is the $j^{th}$ eigenvalue with
only the first term kept in (16), corresponding to infinite
separation $d\rightarrow \infty $. Note that the even and odd
modes are degenerate in the limit $d\rightarrow \infty $,
acounting for the factor $2$ in the last term of
(\ref{EvdWEvenOdd}).

\subsection{Numerical Convergence}
Numerical convergence of the dispersion interaction with respect
to $N$ was also verified. A larger $N$ corresponds to dividing the
integral into more partitions; consequently, as more ``points''
along the wire are sampled, larger $N$ means plasmon modes with
shorter wavelengths are also sampled. The lowest plasmon mode
wavelengths are expected to contribute negligibly to the
dispersion interaction, so for sufficiently large $N$ numerical
convergence of the vdW interaction should be evident. We found $N$
up to $800$ was sufficient for convergence of the dispersion
interaction for wires separated by $D=2A$ at all of the lengths
treated here (up to $80A$). Increasing the internal pressure,
corresponding to using a larger Fermi velocity, gives a more rapid
convergence of the dispersion interaction with respect to $N$ and
a weaker vdW attraction between the two wires. This is because the
pressure term stiffens the short--wavelength modes, making them
less sensitive to the weak interaction between the wires. We also
found that inclusion of the pressure improved the numerical
stability.

The value of $N$ necessary for convergence (to a given fractional
level) increases with wire length and also with wire separation
distance. For comparison of the dispersion energy for conducting
and insulating wires, and in investigating variation of separation
distance for wires much longer than the separation distance, we
employed $N$ sufficiently large to achieve convergence of the vdW
interaction to about three significant figures.

\subsection{Choice of numerical parameters for carbon nanotubes.}

The basic approach described above should apply to any system
where there is an essentially one-dimensional electron gas with
long mean free path, such as a metallic nanowire or the conduction
band of a metallic (n,n) carbon nanotube. The results using
(\ref{DimlessEquMotion}) are dimensionless and therefore universal
in a sense. However the dispersion energy from (\ref
{DimensionedEDisp}) is thereby obtained in units of $\hbar \,
\Omega _{0}$ defined in (\ref{Omrga0}), and this quantity will be
different for different 1D systems. Our dimensionless lengths are
defined in terms of the Coulomb smearing length $A$, which is
taken to be the radius of the quasi-1D electron gas, e.g. the tube
radius in the nanotube case. It is also necessary to choose the
dimensionless input parameters $\beta $, $\omega _{pin}\equiv
\Omega _{pin}/\Omega _{0}$ to suit the particular system.

For the numerical work below we chose parameters roughly
appropriate to the conduction band of a (5,5) single-walled carbon
nanotube. While a metallic nanotube is an excellent metal, it is
not a \emph{free-electron} metal. We therefore first establish a
correspondence between our classical Newton II approach and the
relevant $q\rightarrow 0$ quantal response by considering an
infinitely long wire as follows. A simple model of the rigid
time-oscillatory displacement of the k-space Bloch electron
distribution under an external electric field
$E_{0}\,\exp(iqz-i\omega t)$ gives the bare density-density
response for $q\rightarrow 0$ as
\begin{equation}\label{QuantalChi0}
\chi ^{0}(q,\omega )=v_{F}q^{2}/(\pi \omega ^{2}\hbar )\,.
\end{equation}
Here the Fermi velocity $v_{F}=\hbar ^{-1}\partial \varepsilon
(k)/dk|_{k=k_{F}} $ is the gradient of the 1-dimensional Bloch
energy $\varepsilon (k)$ at the Fermi point $k=k_{F}$. $v_{F}$ can
be calculated by differentiation of the analytic Bloch electron
dispersion for single-walled (n,n) nanotubes given in Eq (4.6) of
Saito and Dresselhaus \cite{PhysPropsCNanotubesSaitoDresselhaus}.
In this way we obtained the Fermi velocity of the conduction bands
(those with $q=n$ and $q=2n$ in the notation of Saito and
Dresselhaus) as $v_{F}=\sqrt{3}a\left| t\right| /(2\hbar)$.  Here
$\left| t\right| =3.03\;eV$ is the hopping parameter and
$a=4.65\,$Bohr radii is the length of the primitive translation
vector of the parent 2D graphene lattice. This gives
$v_{F}=9.9\times 10^{5}m/s$ independent of $n$, a value quite
comparable to Fermi velocities in 3D metals.

The bare response of the infinite wire can also be calculated via the Newton
II approach (used in our numerical treatment outlined above for the case of
finite length), giving
\begin{equation}\label{ClassicalChi0}
\chi _{0}(q,\omega )=n_{0}q^{2}/(m\omega ^{2})
\end{equation}
where $n_{0}$ is the 1D density of the electron gas and $m$ is the
effective classical mass.

By comparing (\ref{ClassicalChi0}) and (\ref{QuantalChi0}) we establish the
equivalence
\[
n_{0}/m=v_{F}/(\pi \hbar )\,.
\]
Thus the energy $\hbar \Omega _{0}$ that scales our vdW energy
predictions (see Eqs (\ref{Omrga0}), (\ref{DimensionedEDisp}) and
(\ref{DimlessEquMotion})) is
\begin{equation}\label{hbaromega}
\hbar \Omega _{0}=\hbar \sqrt{\frac{v_{F}}{\pi \hbar
}\frac{e^{2}}{A^{2}}}= \frac{8.02}{n}\,eV
\end{equation}
where we have used the result
\cite{PhysPropsCNanotubesSaitoDresselhaus} $A=\sqrt{3}na/(2\pi )$
for the radius of a (n,n) carbon nanotube.

Assuming that the characteristic velocity $B$ entering the
pressure term Eq (\ref{pressforce}) is of order $v_{F}$, we find
that the dimensionless pressure parameter $\beta $ appearing in
(\ref{DimlessEquMotion}) is of order unity. We used the value
$\beta =1.38$ throughout.

We also need to choose values for the pinning frequency $\Omega
_{pin}$ that mimic the effect of an insulating energy gap.
Semiconducting carbon nanotubes have electron energy gaps up to
$O(1\;eV)$. \ Assuming that $\hbar \Omega _{pin}$ is of this
order, we obtain from the numbers above a ratio $\omega
_{pin}\equiv \Omega _{pin}/\Omega _{0}=O(n/8)$ for a (n,n)
nanotube. We explored cases with values of $\omega _{pin}$ varying
from $0$ (representing the conduction electrons of a metallic
wire) to $\omega _{pin}=1.73$ (representing strongly
semiconducting electron bands).

\section{Results and Discussion}

\label{resultsanddisc}

\begin{figure}[tbp]
\begin{center}
\includegraphics[width=8cm]{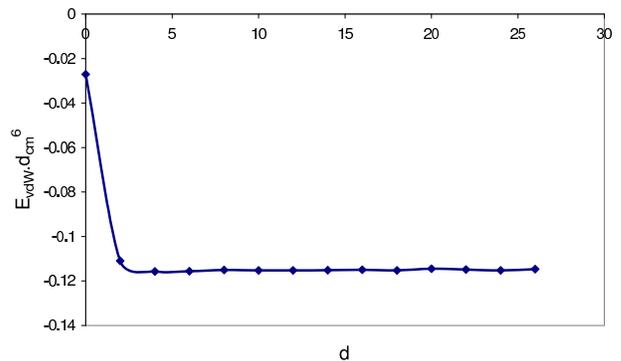}
\end{center}
\caption{Dispersion energy $d_{cm}^6E^{vdW}$ vs. $d$ for
dimensionless length $\ell=2$, pressure parameter $\beta=1.38$,
and zero pinning.} \label{compasdf}
\end{figure}

We first verified that, in the ``pointing" geometry of interest
here, our numerical coupled-plasmon approach yields the expected
dispersion interaction of form $E_{vdW} \sim -C_6(\ell )d^{-6}$
for two wires whose (equal) dimensionless length $\ell $ is much
less than their dimensionless separation $d$ measured between
their ends. In this limit the $d^{-6}$ law can be predicted
analytically by modelling each piece of wire as a polarizable
dipole \cite{JFDIntJChem05,JFDAstJChem01}.  To test this power law
explicitly, in Fig.\ \ref{compasdf} we plot $d_{cm}^6E_{vdW}\,$\
vs $d$, where $ d_{cm}\equiv d+\ell $ is the distance between the
centers of the wires.

\begin{figure}
\begin{center}
\includegraphics[width=8cm]{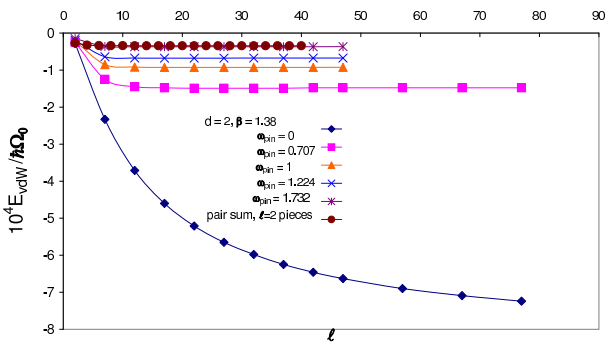}
\end{center}
\caption{Dispersion energy $E_{vdW}$ vs. wire length $\ell$ for
fixed
separation $d=2\,\,$and pressure parameter $\beta =1.38$\\
Diamonds: metallic wire ($\omega _{pin}=0$). Solid squares:\
Semiconducting wire with $\omega _{pin}=0.707$. Solid triangles:
semiconducting wire with $\omega _{pin}=1.0$. Crosses:\
semiconducting wire with $\omega _{pin}=1.224$ . Asterisks:\
semiconducting wire with $\omega _{pin}=1.732$. Solid circles:\
Metallic wire ($\omega _{pin}=0$) that has been cut into pieces of
length $2.$ The unit of energy (on the vertical axis) is $\hbar
\Omega _{0 \text{\thinspace }}$ as defined in Eq (\ref{Omrga0})
and (\ref{hbaromega}). For the conduction band of a (5,5) carbon
nanotube, this energy unit is $1.6\,$eV. } \label{Fig4}
\end{figure}

For each chosen value of $\ell $, the plot settles down to a
constant value $-C_6(\ell )$ for $d>>\ell $, which confirms the
expectation. This basic $d^{-6}$ behavior is observed both for
metallic wires and for semiconducting wires (modelled by adding a
pinning force as described above). \ For a given wire length $\ell
$, the coefficient $C_6(\ell )$ depends considerably on the
pinning frequency $\omega _{pin}$ i.e. metallic wires ($\omega
_{pin}=0$ ) attract more strongly than insulating ones with the
same concentration of van der Waals interacting electrons, but
still with a $d^{-6}$ law for the energy.

Much more interesting is the behavior of the dispersion energy
when the separation $d$ is less than the length $\ell $. Here a
single dipole does not adequately represent the electronic
response of a wire, and consequently no single power law $d^p$
emerges for the energy $E_{vdW}(d)$ because a multipolar expansion
would be required in order to generalize the dipolar argument. We
are therefore out of the ``asymptotic'' region of a distant vdW
interaction, though still not in a region of electronic cloud
overlap. In this regime, a conventional approach would be to sum
(integrate) contributions of form $-C\left| r_1-r_2\right| ^{-6}$
from small segments of the two wires at positions $r_1$ and $r_2$
respectively. This approach predicts that, for fixed $d$, the
dispersion energy for increasing wire length $\ell $ will saturate
rapidly as soon as $\ell $ exceeds $d$, to a value proportional to
$-d^{-4}$. We find that this saturation does occur for insulating
wires, as modelled either with a substantial pinning frequency
$\omega _{pin}$, or by cutting a metallic wire into mutually
insulated pieces shorter than the separation.  It is definitely
NOT\ true for intact metallic wires ($\omega _{pin}=0$) as
modelled by our full numerical coupled-plasmon approach. In the
metallic case the interaction continues to grow with wire length
and still has not saturated for wires many times longer than the
separation. These findings are illustrated in Fig.\ \ref{Fig4}.

\begin{figure}
\begin{center}
\includegraphics[width=8cm]{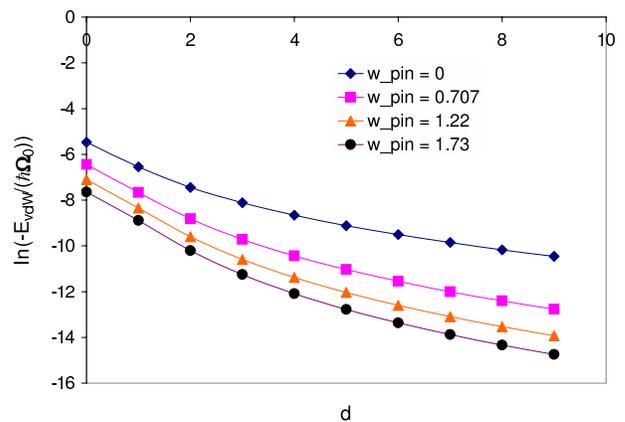}
\end{center}
\caption{Coupled-plasmon interaction at small wire separation
$d$.} \label{Fig5}
\end{figure}

We also investigated the behavior of our coupled-plasmon
interaction right down to zero separation of the wires:\ see Fig.\
\ref{Fig5}. Of course a realistic treatment of this regime would
require the inclusion of covalent and other forces that are not
described in our formalism and that are specific to the particular
quasi-1D system. Furthermore a detailed plasmon treatment even of
the dispersion part of the energy in this regime would require
electronic response functions beyond the long-wavelength
description that we have used. However our results establish the
important result that our method saturates naturally to a finite
value at contact, unlike empirical van der Waals correction
schemes of the $\sum C_{ij}R_{ij}^{-6}$ type, which have to be cut
off in a somewhat arbitrary manner to avoid divergence. Another
significant conclusion from our numbers in this small-separation
regime concerns the relative strength of the dispersion energy for
metallic and nonmetallic cases. Because the analytic results in
various geometries \cite{JDAWARvdwprl06,
vdWParallelWiresChainsChang1971} for the metallic interaction fall
off with a lower power and hence dominate the insulating result at
large distance, one might speculate that the opposite ordering
holds at small separation - i.e. that the insulating interaction
is stronger than the metallic one at small separation. However
such an argument depends on the assumption of the same single
power law for small separations as well as large. This is not the
case, and at small separations $\ d$ a power series rather than a
single power would be required for this type of analysis. In fact
our results show that the metallic attraction is stronger than the
insulating one at all separations, for constant electron number
density.

\section{Summary and conclusions}

We have explored the dispersion interaction between two collinear
one-dimensional electronic systems (``wires'') separated by a
vacuum gap, in the electromagnetically non-retarded regime. We
considered the metallic case and also two models of an
insulating/semiconducting case.  The metallic case was modelled
via free one-dimensional electrons with an electronic pressure
term, but this model was also matched to a quantum band model for
the electrons for the purpose of modelling the conduction electron
bands in carbon nanotubes. For the insulating cases, either a
harmonic pinning force was added to the equations, or a model of
very short, mutually insulated metallic pieces was used, with
rather similar results to the case of a pinning force.  Our
calculations provide significant insight additional to that
previously obtained from asymptotic dispersion energy calculations
\cite{vdWParallelWiresChainsChang1971, SB98QuWells,BS00MeFilms,
JFDIntJChem05, JDAWARvdwprl06} on low-dimensional systems with at
least one infinite spatial dimension.  These previous analytic
calculations were all performed for the asymptotic regime where a
single power law (sometimes multiplied by a logarithm) suffices to
describe the dispersion energy as a function of separation $D$.
These previous works showed that the distant interaction in these
highly anisotropic systems shows major qualitative differences
between the metallic and insulating cases.  Specifically, the
power law is different in the two cases, with the metallic
interaction being stronger in the asymptotic regime studied.

In contrast to those fully asymptotic calculations, we treated
wires of finite length $L$, with a finite separation $D$ between
their ends, a regime where a single power law in $D$ does not
describe the interaction.  Unlike the asymptotic cases, we find
that the interaction at finite separations is a continuous
function of the pinning frequency (i.e. a continuous function of
the electronic energy gap).  Our coupled-plasmon dispersion energy
does however remain a strong function of the gap right down to the
limit of contacting wires, where it remains finite, unlike
asymptotic formulae for the same quantity. In particular we find a
large difference between the attractive dispersion energy of the
most polarizable electron bands of metallic and semiconducting
carbon nanotubes.  The dispersion energy was found to decrease
monotonically with energy gap at all values of $L$ and $D$ that we
studied.

We also found that a conventional pairwise summation of terms of
form $-CR^{-6}$ is not an adequate description of metallic wires,
with a major discrepancy between this model and our
coupled-plasmon results. Specifically, the coupled-plasmon energy,
while similar to the summation approach for very short wires,
continues to grow with length $L$ at fixed separation $D$, while
the $-\Sigma CR^{-6}$ energy saturates rapidly to an $L\rightarrow
\infty$ value that is much smaller than the coupled-plasmon
energy.

A more detailed microscopic model, including a detailed orbital
description with Pauli repulsion and non-dispersive bonding
forces, will be required to describe the cohesive forces at
distances where the electronic clouds overlap. The present results
suggest, however, that long-wavelength coupled charge fluctuations
will still enhance the van der Waals force at these distances,
necessitating a highly nonlocal model of the Coulomb screening
even in this limit. Theories that yield a $\sum R^{-6}$
asymptotics cannot be relied upon to achieve this.

\bibliographystyle{apsrev}

\end{document}